%% file: v6.tex
\documentclass[journal]{IEEEtran}
\makeatletter
\def\endthebibliography{%
	\def\@noitemerr{\@latex@warning{Empty `thebibliography' environment}}%
	\endlist
}
\makeatother

\usepackage{cite}

\ifCLASSINFOpdf
   \usepackage[pdftex]{graphicx}

\else
\fi
\usepackage{amsmath}
\usepackage{array}

\usepackage{tabularx}

\ifCLASSOPTIONcompsoc
  \usepackage[caption=false,font=normalsize,labelfont=sf,textfont=sf]{subfig}
\else
  \usepackage[caption=false,font=footnotesize]{subfig}
\fi
 \usepackage{dblfloatfix}
\usepackage{mathtools}

\usepackage{xtab}

\begin{document}
%
\title{Supervised load identification of 18 fixed-speed motors based on their turn-on transient current}
%
%
%


\author{~Christian~Gebbe
	,~Adil~Bashir
	~and~Thomas~Neuh\"auser
}

%
%

\markboth{XXX,~Vol.~XXX, No.~XXX, XXXX}%
{Shell \MakeLowercase{\textit{et al.}}: Bare Demo of IEEEtran.cls for IEEE Journals}
%



\maketitle

\begin{abstract}
Several studies have already shown that the transient current can be successfully used to identify electric loads. However, most of the proposed methods were validated using only a handful of loads whose electric properties often differed significantly from each other. Therefore, a much more challenging empirical study was carried out here using 18 different fixed-speed motors in a real work environment. A further difficulty was introduced by normalizing the current amplitude to a value of 1~A during steady state. It is shown that a classifier can distinguish these 18 motors even under those conditions with an f1-score of 97.7~\% after a supervised training period. In addition to that it was tested whether the mechanical output type (pump, fan or compressor) of the fixed-speed motors could be inferred using only the transient current. However, the classification results indicate that this is not possible.
\end{abstract}

\begin{IEEEkeywords}
Load identification, Nonintrusive load monitoring, fixed-speed motor, transient current, classification, current harmonics
\end{IEEEkeywords}

%
\IEEEpeerreviewmaketitle

\section{Introduction}
Load identification is an important step of nonintrusive load monitoring (NILM) \cite{zoha2012non}. The term "load identification" refers to classification methods which map input data, usually electric properties derived from the current and/or voltage, to a certain load type \cite{du2010review}.

This classification problem has already been tackled in several papers \cite{sultanem1991using,leeb1993development,leeb1993conjoint,norford1996non,khan1997multiprocessor,leeb1995transient,cole1998data,lee2003electric,cox2006transient,srinivasan2006neural,patel2007flick,chang2007load,yang2007design,lam2007novel,akbar2007modified,chang2008load,katsukura2009life,kushiro2009non,shaw2008nonintrusive,proper2009field,chang2010load,chang2012new,shrestha2009dynamic,ruzzelli2010real,chang2012non,lin2012application,tsai2012modern,bouhouras2012load,chang2013particle,fernandes2013load,paradiso2013ann,he2013incorporating,chang2014power,chang2010load,lin2014development,bier2014disaggregation,vsira2015system} and different features and classifiers have been proposed (see section~\ref{sec:attach_table_review}). However, most of the proposed methods were validated using only a handful of loads, whose electric properties often differed significantly from each other (see section~\ref{sec:attach_table_review}). We asked ourselves: Is a classification still possible
\begin{itemize}
	\item if the loads are very similar to each other,
	\item if the loads are normalized with respect to their current draw during steady state,
	\item if there are much more than only a handful of loads?
\end{itemize}
Therefore, we conducted an empirical study using 18 different fixed-speed motors in a real working environment. Thereby, we directly address an issue raised by \cite{zeifman2011nonintrusive}, who state that "load identification performance using STFT [short time Fourier transform] has not been characterized for many practical scenarios".

In addition to that we wanted to know whether the type of mechanical load (i.e., a pump, a fan or a compressor) could be inferred automatically after a supervised training. The difference between these three mechanical
loads is that a pump transfers a liquid from a region of low pressure to a region
of high pressure, whereas a fan moves a gas with adding negligible amounts
of pressure to it, and a compressor (and a blower) raises the pressure of a gas significantly \cite{marchildon2014pumps}. This analysis was motivated by the statement that the "transient behavior of a typical load is intimately related to the physical task that the load performs" \cite{leeb1993development}. Moreover, Sultanem \cite{sultanem1991using}, one of the earliest publication in this area, already proposed to distinguish pump-operated appliances from other motor-driven appliances. Since fixed-speed motors represent one of the most frequently occurring loads in industry, a classifier which could infer its mechanical load by measuring its current would be of great benefit.

The rest of the paper is structured as follows: Section~\ref{sec:load_identification_theory} describes the load identification problem in more depth from a theoretical point of view. After that, the data and methods used in this case study are reported in detail in section~\ref{sec:data_methods}. The results of the classification are reported in section~\ref{sec:results}. Finally, the paper is briefly summarized in section~\ref{sec:summary}.

\section{Load identification theory}
\label{sec:load_identification_theory}
As mentioned above, load identification refers to classification methods which map input data to a certain load type \cite{du2010review}. In this section, the possible load types and the kind of classification problem shall be explained in more depth.

\subsection{Possible load types}
In a lot of papers it is not clearly specified, which load types exist. Several options are conceivable and have been (implicitly) proposed in papers:
\begin{itemize}
	\item One type for each operational status for each (physically different) load
	\item One type for each (physically different) load
	\item One type for each device model (e.g. dishwasher of company A, model X)
	\item One type for each device type (e.g. dishwasher)
	\item One type for each group of devices (e.g. linear loads)
\end{itemize}
For example, \cite{bouhouras2012load} partitions a hair dryer and a electric heater into their different operational statuses (e.g. full power, half power and standby). Opposed to that, \cite{bier2014disaggregation} uses two different hair dryers in his case study but does not distinguish them further by their operational status. In contrast to that, \cite{paradiso2013ann} groups all washing machines, dishwashers, etc. into one load type. Similarly, \cite{leeb1995transient} groups two different induction motors into one load type. Last, \cite{vsira2015system} mentions that the signatures from an electric heater and an electric iron appear very similar, indicating that grouping them together might be beneficial for increasing the classification accuracy. Similarly, \cite{sultanem1991using} proposes a load category "resistive appliances".

In this paper, we only want to recommend to clearly specify the load types, but we shall not evaluate which load types are the most sensible choice. Such an evaluation will certainly need to take the final purpose of the classification into account which may differ from application to application. Moreover, the choice of load types might also be influenced by the choice of features: While \cite{srinivasan2006neural} reports that devices of the same model exhibit very similar features, \cite{gupta2010electrisense} noticed a discernible variation in the voltage noise of four compact fluorescent lights of the same model, purchased as a packaged bundle, by which they could be distinguished.

\subsection{Kind of classification problem}
In the simplest case, the current and/or voltage of one load is measured, relevant features are extracted and exactly one load type is predicted. Since the classifier needs to know the load types a priori, the classifier is provided with samples of each load type during training. Thus, we are dealing with a supervised classification.

Often, the current and/or voltage of an aggregated load comprising many individual loads is measured. In these cases, the switching events are usually detected at first \cite{du2010review} and it is assumed that each switching event is caused by exactly one load. For each switching event, features are extracted by using some kind of disaggregation method and one load type is predicted. Thus, the problem simplifies to a supervised classification problem again. However, switching events, in particular turn-on events, may also overlap. \cite{leeb1993development} solve this problem by only considering a section of a turn-on event, which they then assume to be caused by only one load.

In all of the descriptions above, only one class is inferred for a particular input. In contrast to that, papers using artificial neural networks as classifiers, e.g. \cite{srinivasan2006neural,chang2007load, chang2014power}, often have one binary output neuron for each possible load type. These output neurons represent whether a load of a certain type is present ("on") in the measured aggregate load. Thereby, the presence of multiple load types can be predicted at once. Such classifiers are often trained with all possible load type combinations. Since there are $2^N$ possible combinations assuming $N$ different load types, this approach does not scale well for large numbers of $N$. However, posing the classification problem in this way makes sense if no event detector is used or if overlapping events are likely. 

Last, some nonintrusive load algorithms \cite{suzuki2008nonintrusive, egarter2013evolving, kim2011unsupervised, parson2014unsupervised} pose the disaggregation problem as an optimization problem by trying to match the measured aggregate power demand as closely as possible and modeling appliances as e.g. Markov models. While such algorithms also implicitly estimate which loads are active at which times, the structure of such algorithms deviates significantly from  the supervised classification methods described above. Thus, such algorithms shall not be further analyzed in this paper.

\section{Data and methods}
\label{sec:data_methods}

In the following, information about the dataset, the data acquisition methods, the preprocessing methods, the feature extraction methods and the classification method is presented.

\subsection{Dataset}
The dataset includes the electric measurements from 18 different fixed-speed motors (see table~\ref{tab:data_motors}). All motors are part of machines which are used in manufacturing plants and were located at the workshop of Fraunhofer IGCV. For each motor, the type of mechanical load (pump, compressor, fan or other) was determined. No more specifications about the motors was obtained due to often limited documentation and difficulties with disassembling the machine to analyze the motor from close range. However, it is assumed that all motors differ from each other in terms of the device model, since the machines come from different manufacturers and the motors serve different purposes.

\begin{table}[htbp]
	\caption{Overview of analyzed fixed-speed motors}
	\begin{tabular}{p{3.5cm}p{0.5cm}p{0.5cm}p{0.5cm}p{0.5cm}p{0.8cm}}
		\hline
		Machine & Pump & Com-pressor & Fan & Other mech. Load & Number of turn-on events \\ \hline
		Thermoform machine - vacuum pump &  & 1 &  &  & 14 \\ \hline
		Thermoform machine - conveyor belt &  &  &  & 1 & 12 \\ \hline
		Thermoform machine - motor for winding &  &  &  & 1 & 21 \\ \hline
		Thermoform machine - cooling unit & 1 &  &  &  & 12 \\ \hline
		Thermoform machine - cooling unit &  & 1 &  &  & 8 \\ \hline
		Thermoform machine - cooling unit &  &  & 1 &  & 9 \\ \hline
		Heated washing basin & 1 &  &  &  & 27 \\ \hline
		Milling machine - cooling lubricant pump & 1 &  &  &  & 8 \\ \hline
		Motor driving a generator &  &  &  & 1 & 10 \\ \hline
		Pedestal fan &  &  & 1 &  & 28 \\ \hline
		Cooling unit for selective laser sintering machine 1 & 1 &  &  &  & 15 \\ \hline
		Cooling unit for selective laser sintering machine 1 &  & 1 &  &  & 38 \\ \hline
		Cooling unit for selective laser sintering machine 1 &  &  & 1 &  & 39 \\ \hline
		Cooling unit for selective laser sintering machine 2 & 1 &  &  &  & 47 \\ \hline
		Cooling unit for selective laser sintering machine 2 &  & 1 &  &  & 46 \\ \hline
		Ultrasonic cleaner - fan &  &  & 1 &  & 14 \\ \hline
		Ultrasonic cleaner - fan for drying &  &  & 1 &  & 18 \\ \hline
		Ultrasonic cleaner - circulation pump & 1 &  &  &  & 10 \\ \hline
		SUM & 6 & 4 & 5 & 3 & 376 \\ \hline
	\end{tabular}
	\label{tab:data_motors}
\end{table}

\subsection{Data Acquisition}
The current of each motor was measured by placing a current clamp around the cables leading to the motor, usually in the control cabinet. Similarly, the voltage was obtained by placing a magnetic voltage probe to live metal parts in the control cabinet. In case of three-phase motors, only one (arbitrary) phase was measured.

\begin{figure} [htb]
	\centering
	\includegraphics[width=3.5in]{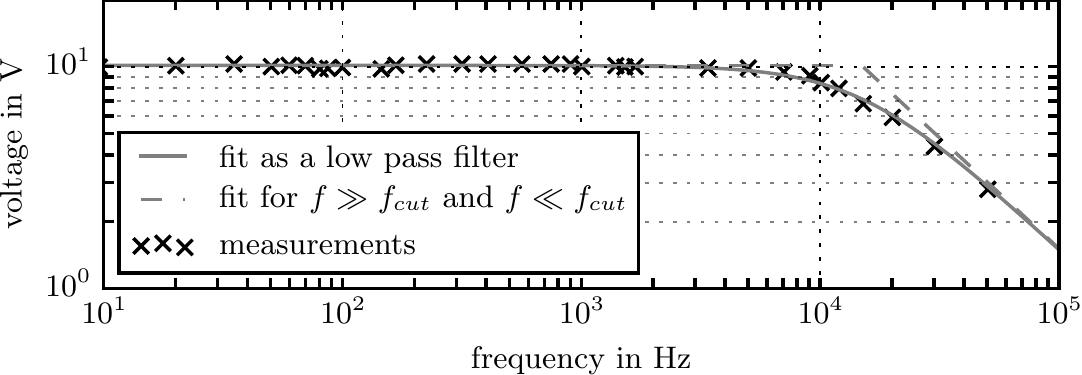}
	\caption{Experimentally determined frequency response of the current clamps, which output a voltage signal}
	\label{fig:data_current_clamps}
\end{figure}

As current clamps, the model WZ12B from the company GMC-I Messtechnik GmbH were used, for which a measurement error of $\epsilon= \pm1.5 \cdot i \pm 1mA$ is stated. Even though they are only specifed for a frequency range of 45 Hz to 500 Hz, it was experimentally determined by both the author and the manufacturer that they measure a current with a frequency of 10 kHz with an attenuation factor of only a few percent (see Fig.~\ref{fig:data_current_clamps}).

As data acquisition device, the model DS-NET from the company DEWESoft GmbH was used. Its V4-HV module was used for recording the voltage, and its V8 module to record the voltage output of the current clamps. Both were measured with a sample frequency of 10~kHz.

In this paper, we are intentionally measuring the current of the motor directly (similar to previous papers, see section~\ref{sec:attach_table_review}). In contrast to that, NILM devices only measure the current of an aggregate load. As described above, loads are then identified by detecting switching events and extracting features for each switching event which includes some kind of disaggregation method. Thus, load identifcation using the aggregate current actually comprises two problems: A disaggregation and a classification problem. Here, we are only tackling the classification problem. The reason is as follows: If the load identification using the aggregate current yields poor results, it is not clear whether it is due to the disaggregation or the classification method. On the other hand, if the load identification using the motor current yields poor results, it is clear that the issue is the classification method. 

\subsection{Preprocessing methods}
\label{subsec:methods_preprocess}
\begin{figure} [htb]
	\centering
	\includegraphics[width=3.5in]{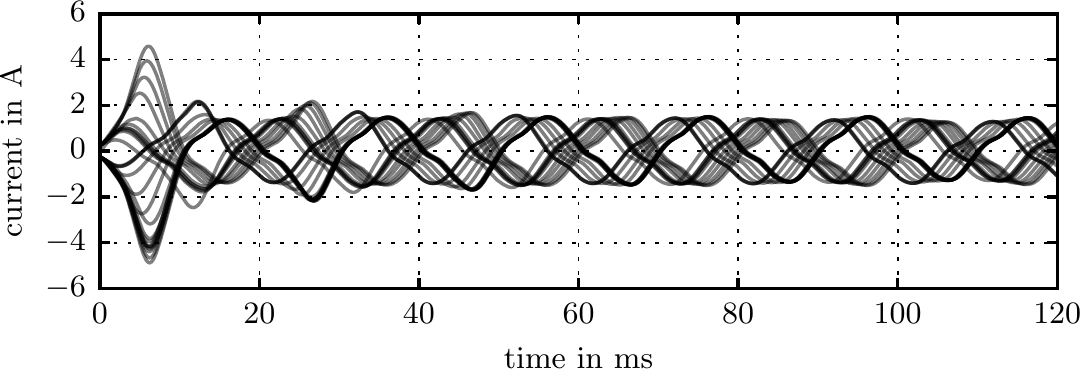}
	\\
	\vspace{0.2cm}
	\includegraphics[width=3.5in]{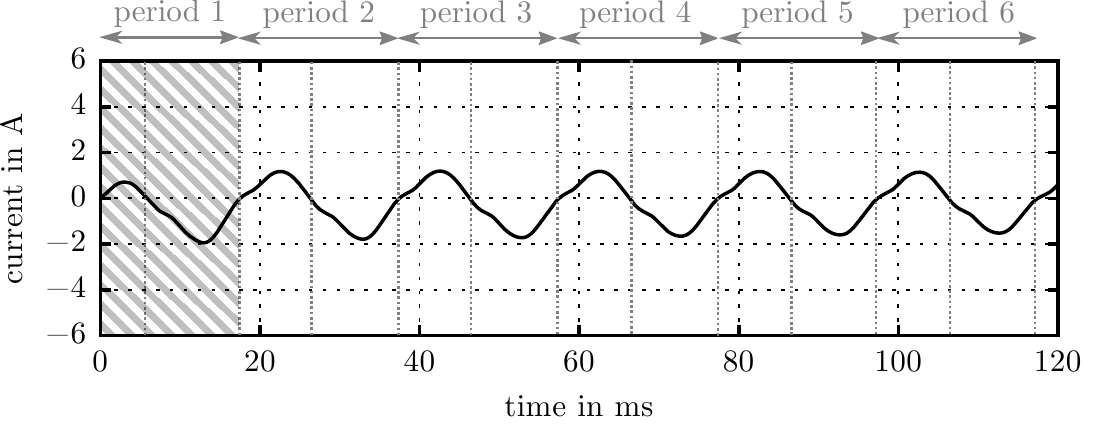}
	\includegraphics[width=3.5in]{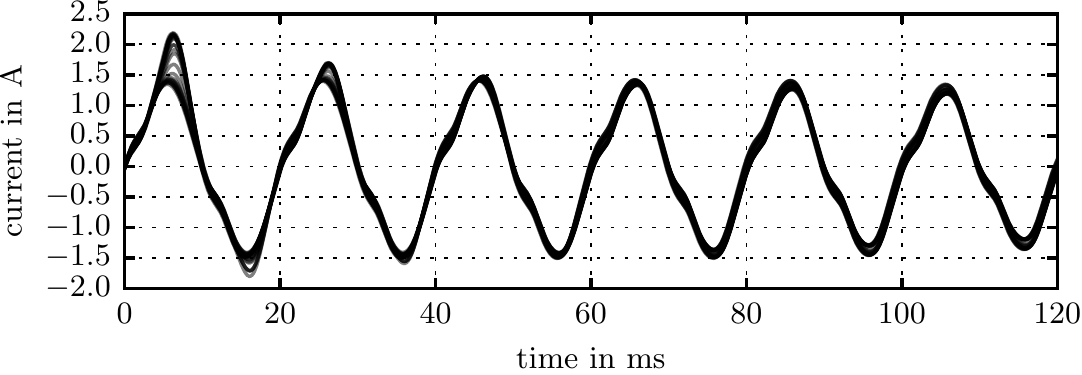}
	\caption{Transient current of 27 turn-on events of the heated washing basin before (top) and after (bottom) performing the preprocessing (middle).}
	\label{fig:data_preprocess}
\end{figure}

The current measurements of each motor were divided by a constant factor so that the current peaks during steady state were at approximately $\pm1~A$. Then, the turn-on events of each motor were detected using a simple threshold algorithm. The turn-on currents of a motor usually varied significantly, most likely due to a different phase angle at the time of turning-on (see Fig.~\ref{fig:data_preprocess} top).

In order to mitigate this influence, the current was split into periods using a zero-crossing approach and the first period corresponding to the first approximately 20~ms was omitted (see Fig.~\ref{fig:data_preprocess} middle). Moreover, the current was potentially multiplied by $(-1)$ so that the first current peak is positive. After this preprocessing, all turn-on currents of a fixed-speed motor looked much more similar (see Fig.~\ref{fig:data_preprocess} bottom). Based on the preprocessed current $i$ and the measured voltage $u$, the instantaneous power $p_{inst}$ was calculated as $p_{inst}(t)=i(t) \cdot u(t)$. 

\subsection{Extracted features}
\label{subsec:methods_features}
For each turn-on event, several features were calculated. Therefore, only the time scale of the second to sixth period of the current after the turn-on event was considered, which corresponds to the first approximately 20-120~ms after a turn-on event (see Fig.~\ref{fig:data_preprocess} middle). While later times would probably provide useful features, too, there is higher chance that another turn-on event occurs during that time period. Because of that, the duration of the transient was \emph{not} used as a feature, since it can be much longer than 120~ms. The following features were calculated (see summary in Table~\ref{tab:06_class_features}):

\paragraph*{Exponential decay constant}

An exponential function of the form \mbox{$i_{peak}(t) = i_{max} \cdot e^{-\lambda t} + i_{steady}$} was fitted to the current peaks and the parameter $\lambda$ was used as a feature. More exactly, this function was fitted to four different signals resulting in four different features: The five peak maxima of the current, the five peak minima of the current, the ten peak maxima of the absolute value of the current and the ten maxima of the instantaneous power $p_{inst}$.

\paragraph*{Linear slope constant}

Analogous to calculating the exponential decay constant, a linear slope constant $m$ was calculated by fitting a function of the form \mbox{$i_{peak}(t) = i_{max} \cdot m + i_{steady}$}. It was also calculated for the four signals specified above.

\paragraph*{Current peak values}

Another set of features were the peak values of the current in the second to sixth period. In total, five current peak maxima values and five current peak minima values were used as a feature. If we would not have normalized the current during preprocessing, such current peaks would represent absolute values. However, one might prefer using relative values to using absolute values in order to classify identical device types equally, independent of their size. Therefore, the absolute current peak values have to be divided with other absolute values. In this paper, the transient current peak values are implicitly divided by the steady state current peak values during the normalization in preprocessing (see subsection~\ref{subsec:methods_preprocess}). Alternatively, the transient current peak values can be divided by themselves. This approach was also tested. As divisor, the maximum peak value of the third period was chosen, because this particular value varied, on average, the least for all motors compared to the other current peaks in the second to sixth period. Concluding, both the "absolute" (only implicit division with steady state current amplitude) and the "relative" (explicit division with transient current peak value) current peak value were used as features.

\paragraph*{Energy}

Inspired by \cite{chang2007load}, the energy of the transient defined as \mbox{$e(t_0, \Delta_t) = \int_{t_0}^{t_0+\Delta_t} p_{inst}(t) dt$} was used as a feature. More precisely, the energy of each period from the second to sixth period was used (five values) and the \emph{sum} of the energy ranging from the second to either the second, third, fourth, fifth or sixth period (five more values). Since these values represent absolute values again, in addition to the absolute values relative values were calculated by dividing the energy in each period by the energy of the sixth period. The sixth period was chosen, because it varied, on average, the least for all motors compared to the energies of the four periods.

\paragraph*{Current harmonics}

Similar to \cite{leeb1995transient}, the magnitude of the first 20 current harmonics was calculated using a fast Fourier transform. The magnitudes were divided by the magnitude of the first harmonic, resulting in relative features. The fast Fourier transform was applied five times for each of the second to sixth periods, resulting in a total of 100 features. Moreover, the total harmonic distortion was calculated for each period via $THD=(|I_2| + |I_3| + \dots + |I_{20}|)/|I_1|$, where $|I_n|$ represents the magnitude of the n'th current harmonic. This resulted in five more features.

\paragraph*{Presence of additional local extrema and inflection points}

In a perfect sinus signal, there is only one local extrema in the middle of each half-period. However, some currents exhibited multiple local extrema in half-periods. This phenomenon was captured with one feature per half-period which took the value one whenever multiple local extrema appeared and otherwise zero. Similarly, an inflection point, at which the curvature of a signal changes, only appears for a perfect sinus when its value approaches zero. In some half-periods additional inflection points were observed though and in these cases a corresponding feature adopted the value one and otherwise zero.

\begin{table}[htb]
	\centering
	\caption{List of all features calculated in the case study}
	\begin{tabular}{ll}
		\hline
		Feature category & Number of features in category \\
		\hline
		Exponential decay constant & 4 \\
		Linear slope constant & 4 \\
		Peak values - absolute and relative & 20 \\
		Energy - absolute and relative & 10 \\
		Energy sum - absolute and relative & 10 \\
		Magnitude of current harmonics & 100 \\
		Total harmonic distortion of harmonics & 5 \\
		Presence of local extrema & 10 \\
		Presence of inflection points & 10 \\
		\textbf{Sum}   & \textbf{173} \\
		\hline
	\end{tabular}%
	\label{tab:06_class_features}%
\end{table}%

\subsection{Classifier for distinguishing motors}
\label{subsec:classifier_distinguish}

The actual classification can be subdivided into the following five substeps:
\begin{itemize}
	\item Prepare the data set in order to facilitate the classification process
	\item Split the data set into training and test set
	\item Choose features
	\item Choose a classifier
	\item Train a classifier 
	\item Choose a classification accuracy measure
\end{itemize}

\paragraph*{Prepare data set}
Before the classification, all features were rescaled to $[0,1]$. Otherwise, features with a high magnitude are likely to dominate the classification process. There are 18 labels for each (physically different) fixed-speed motor.

\paragraph*{Split data set} 
In order to train and evaluate the classifier, the data set is split into a training set for training the classifier and a test set for evaluating the classifier. This is performed using the stratified k-fold cross validation technique with $k=8$ \cite{refaeilzadeh2009cross}. One weakness of the data set is that the number of turn-on events differs among motors. In machine learning terms the data set is said to be imbalanced. Thus, the classifier would prioritize the correct classification of those motors that have the highest number of turn-on events while it should consider all motors as equally important. The equal treatment of classes can be achieved by modifying the cost function during the training procedure. Therefore, in scikit-learn the parameter class weight was set to 'balanced'.

\paragraph*{Choose features} 
One could train a classifier using all the 173 features defined in the previous subsection. However, it would be desirable to use significantly fewer features to make the classifier simpler and also avoid overfitting. To this end, we followed the following strategy: At first, train a classifier with each of the 173 features (only one feature). Determine the accuracy of each classifier. Pick the "winning" feature which yielded the highest classification accuracy. Then, train a classifier with two features. Limit the $173^2=29929$ possible feature combinations to only $173$ possible combinations by fixing one feature to the "winning" feature of the previous run. Again, train $173$ different classifiers and pick the "winning" feature of the second run. Follow this approach of adding one more "variable" feature for each run and keeping the "winning" features of the previous runs. Thereby, we tested classifiers that use 1-15 features. For a given number of features, only the classifier with the "winning" features is reported in the following.

It is clear that this approach will yield worse results than trying all possible combinations for a fixed number of features. However, for a classifier with 15 features, there are approximately $173^{15}=3.7\cdot10 ^{33}$ possible feature combinations. Trying all these out by brute force would take too much time.

\paragraph*{Choose classifier} 
Two of the most often employed supervised classification techniques are neural networks and support vector machines. While neural networks tend to require a large number of samples \cite{gebbe2017feature}, support vector machines 'are well suited to deal with learning tasks where the number of features is large with respect to the number of training instances' \cite[p.~13]{maglogiannis2007emerging}. Since this condition was true for this case study (the number of potential features was approximately half as large as the number of the samples), the support vector machine was chosen. It was implemented using the function 'sklearn.svm.SVC' from the python module  scikit-learn, version 0.19.2 \cite{pedregosa2011scikit}. One of the parameters to further specify the support vector machine classifier is the choice of the kernel. Here, three kernels were tested: a linear kernel, a polynomial kernel with three degrees and a radial basis function kernel. All other parameters of the function were left to their default values.

\paragraph*{Train classifier} 
Training the classifier means fitting the parameters of the mathematical model of the support vector machine to the training data such that the classification accuracy is maximized. To this end, scikit-learn provided the function ‘fit’. 

\paragraph*{Choose accuracy measure} 
The classification accuracy of the trained classifier was evaluated using the macro-averaged f1-score. This metric is more appropriate than the default accuracy measure (fraction of correct classifications) in case of imbalanced data sets.

\subsection{Classifier for predicting the mechanical output}
\label{subsec:classifier_predict_mech}
The classifier for predicting the mechanical output was the same as the one used for distinguishing the 18 different fixed-speed motors with the following deviations:
\begin{itemize}
	\item The three motors whose mechanical output was neither a pump, a compressor nor a fan, were omitted.
	\item Instead of the 18 labels for each motor, only three labels were used representing the three different mechanical outputs: pump, compressor and fan
	\item The number of turn-on evens per mechanical output was not equal. Therefore, the parameter class weight was set to 'balanced' again.
	\item The number of turn-on events per motor was not equal. To ensure equal importance, the number of turn-on events per motor was reduced to eight through omission. Thereby, only eight turn-on events from each motor were used.
	\item Instead of the default stratified cross-validation we employed a custom cross-validation strategy. The reason is that the default stratified cross-validation folds would likely contain one turn-on event from  each motor. In this way, the classifier could learn to predict the three different mechanical outputs by distinguishing the (18-3=) 15 different motors. Instead, we want the classifier to generalize the characteristics of a mechanical output. Therefore, the data set was split such that the test set includes all turn-on events from one motor from each mechanical output (i.e., the test data includes three motors and a total of 24 turn-on events). All residual turn-on events form the training set. This splitting strategy ensures that the motors in the test data set are new to the classifier. There are $6\cdot4\cdot5=120$ possible combinations to split the dataset in this way (see table~\ref{tab:data_motors}). For each of these combinations, a classifier was trained and its accuracy (\mbox{f1-score}) was evaluated. The final \mbox{f1-score} was determined as the mean of these \mbox{f1-scores}.
\end{itemize}

\section{Results}
\label{sec:results}

First, the results of the classifier for distinguishing motors are presented in subsection~\ref{subsec:result_distinguish}. Then, the results of the classifier for inferring the mechanical output type are discussed in subsection~\ref{subsec:result_distinguish}. 

\subsection{Distinguishing motors}
\label{subsec:result_distinguish}

\begin{figure} [htb]
	\centering
	\includegraphics[width=3.5in]{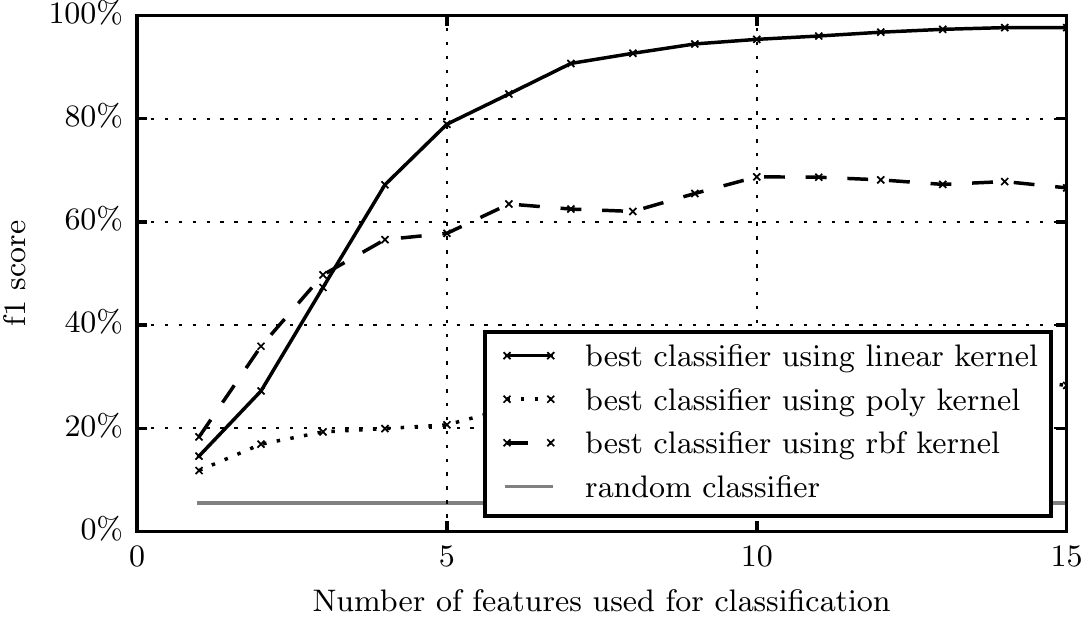}
	\caption{classification results for distinguishing the 18 fixed-speed motors}
	\label{fig:result_motor_plot}
\end{figure}

The classification results for distinguishing the 18 different fixed-speed motors are presented in Fig.~\ref{fig:result_motor_plot}. Among the three different SVM kernels, the classifier using the linear kernel achieved the highest f1-score. It may be surprising that the classifier using the linear kernel performed better than the one with the polynomial kernel, because the linear kernel is just a subset of the polynomial one. However, we believe that the use of the polynomial kernel increased the degree of freedom of the model too much given the limited amount of training data.

The f1-score of the classifier using the linear kernel depends significantly on the number of features used for classification. It is less than 20~\% when only one feature is used, approximately 80~\% when five features are used and 97.7~\% when 14 or 15 features are used. This f1-score is significantly higher than the expected f1-score of $1/18=5.6~\%$ for a randomly guessing classifier.

\begin{table}[htbp]
	\caption{"Winning" feature for each run}
	\begin{tabular}{p{1.3cm}p{5.5cm}r}
		\hline
		number of features & Additional "winning" feature & f1-score \\ \hline
		1 & Magnitude of 2nd harmonic in period 2 & 14.6~\% \\ \hline
		2 & Linear slope constant using maxima of absolute current & 27.2~\% \\ \hline
		3 & Magnitude of 3rd harmonic in period 5 & 47.3~\% \\ \hline
		4 & Absolute peak value of current minimum in period 4 & 67.2~\% \\ \hline
		5 & Magnitude of 7th harmonic in period 5 & 78.9~\% \\ \hline
		6 & Magnitude of 9th harmonic in period 3 & 84.8~\% \\ \hline
		7 & Magnitude of 17th harmonic in period 3 & 90.7~\% \\ \hline
		8 & Relative peak value of current maximum in period 5 & 92.7~\% \\ \hline
		9 & Magnitude of 6th harmonic in period 3 & 94.5~\% \\ \hline
		10 & Magnitude of 3th harmonic in period 6 & 95.4~\% \\ \hline
		11 & Magnitude of 9th harmonic in period 5 & 96.0~\% \\ \hline
		12 & Magnitude of 11th harmonic in period 5 & 96.8~\% \\ \hline
		13 & Magnitude of 11th harmonic in period 6 & 97.3~\% \\ \hline
		14 & Magnitude of 17th harmonic in period 4 & 97.7~\% \\ \hline
		15 & Linear slope constant using current minima & 97.7~\% \\ \hline
	\end{tabular}
	\label{tab:result_motor_features}
\end{table}

The “winning” features for each run are shown in table~\ref{tab:result_motor_features}. Most of these feature represent the magnitudes of certain current harmonics. This might not be too surprising, as 100 out of 173 features represent current harmonic magnitudes. However, we expected that the total harmonic distortion of a period would be a more helpful feature than a rather specific current harmonic. Moreover it was surprising to us that the “winning” feature when using only one feature was the second harmonic in the period~2, an \emph{even} harmonic, whose magnitude is usually negligible compared to any odd harmonic. Apart from the current harmonics, two features repesenting the current peak values directly as well as two linear slope constants could also be found among the “winning” features. In contrast to that, all four exponential decay constants were found to be one of the most unhelpful features. They exhibited the highest average coefficient of variation, calculated for all the turn-on events of one motor, among all the 173 features.

\begin{figure} [htb]
	\centering
	\includegraphics[width=3.5in]{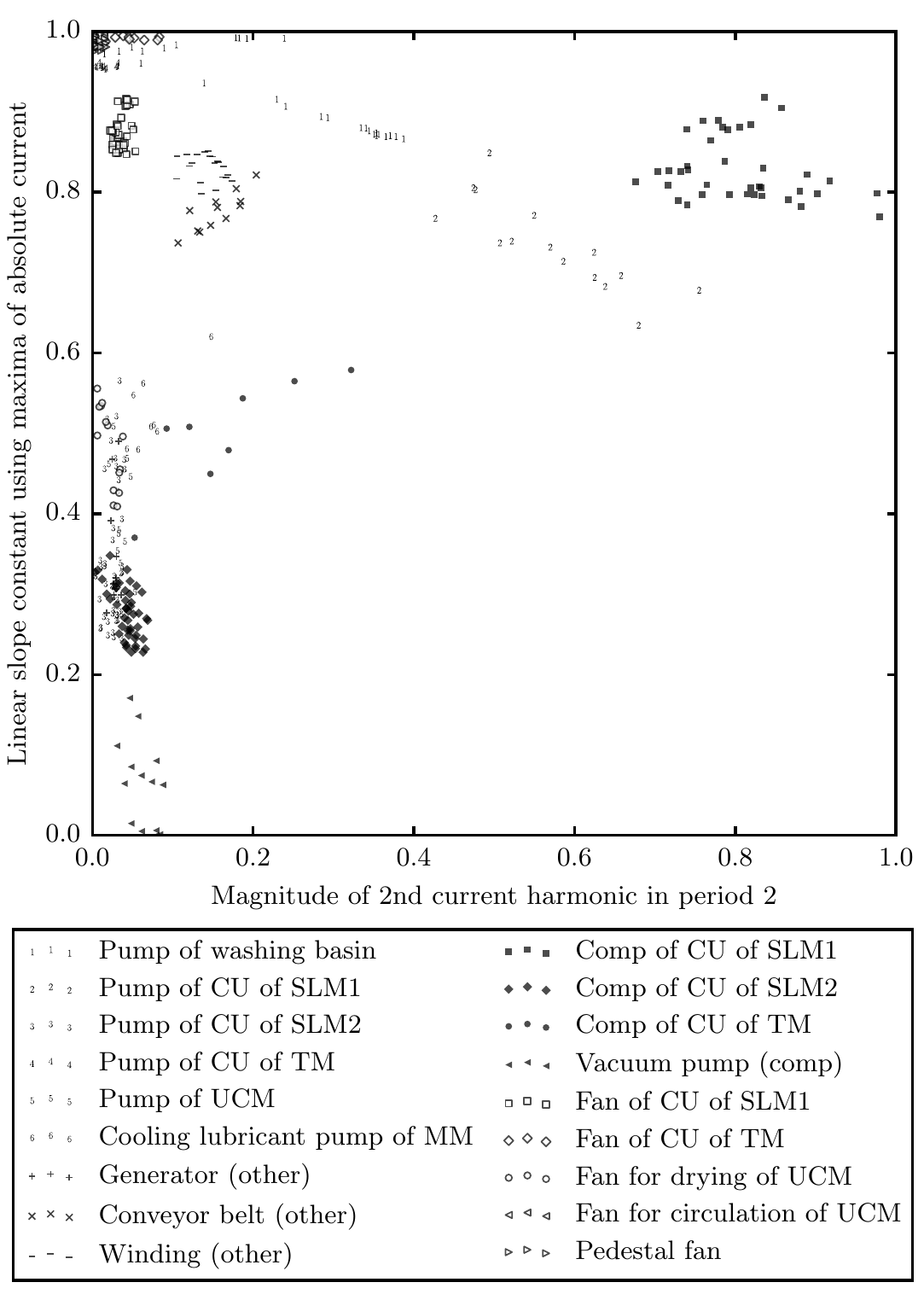}
	\caption{Scatter plot of all turn-on events. CU=cooling unit,
SLM=selective laser sintering machine, TM= thermoform machine, MM=milling machine, UCM=ultrasonic cleaning machine.}
	\label{fig:result_motor_scatter}
\end{figure}

A visual representation of the data is possible by plotting the turn-on events in a two-dimensional scatter plot along the axis of the first two “winning” features (see Fig.~\ref{fig:result_motor_scatter}). It can be seen, that most turn-on events belonging to the same motor are grouped closely together while turn-on events of different motors are separated from each other. This observation is in alignment with the good classification results.

\subsection{Predicting mechanical output of motors}
\label{subsec:result_predict}

\begin{figure} [htb]
	\centering
	\includegraphics[width=3.5in]{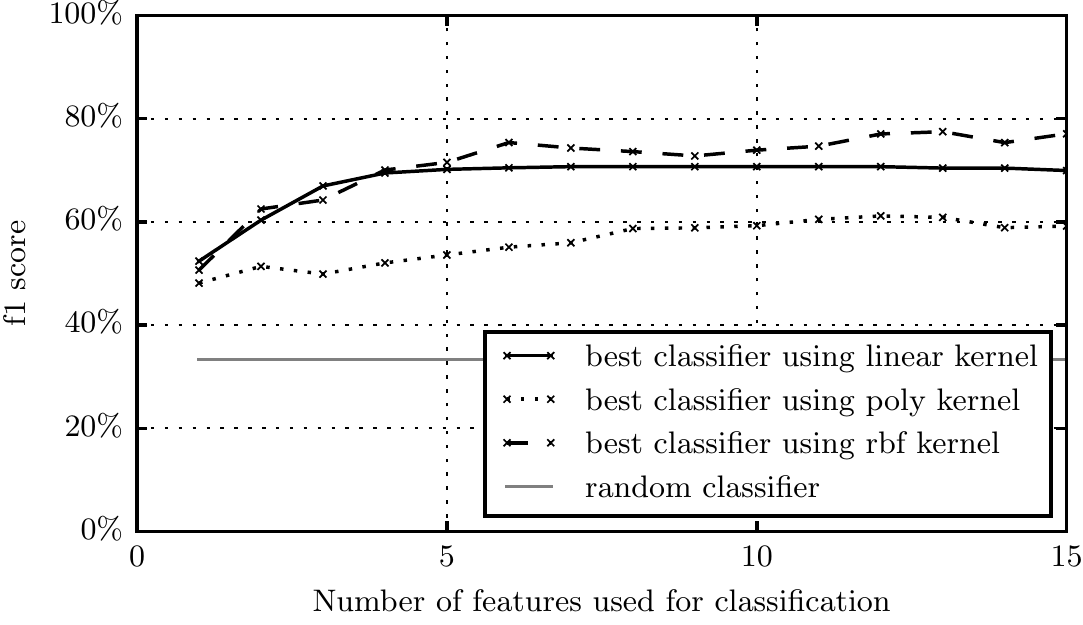}
	\caption{classification results for inferring the mechanical output type}
	\label{fig:result_load_plot}
\end{figure}

The classification results for inferring the mechanical output of a motor are shown in Fig.~\ref{fig:result_load_plot}. Similar to Fig.~\ref{fig:result_motor_plot}, the classifier using the linear kernel outperforms the one with the polynomial kernel. While the classifier using the radial basis function kernel achieves slightly higher f1-scores, its f1-score sometimes decreases if the number of features is decreased which may be sign of a too complex model for the very limited amount of training data. Therefore, we will analyze the classifier using the linear kernel in the following.

Its resulting f1-score depends on the number of features: Using only one feature (relative peak value of current minimum in period 6) it achieves an f1-score of already 52.4~\%. Using five features, an f1-score of 70.7~\% is reached, but it does not increase more using any additional features. While these values seem high, they have to be compared to a randomly guessing classifier whose f1-score is $1/3=33~\%$. While there is an increase in classification accuracy, it does not seem significant.

Thus, it seems that no classifier could be found that performs significantly better than a randomly guessing one. This may be due to the very limited amount of training data which comprises only $4-6$ motors for each type of mechanical load. Potentially, the common characteristics determined by the mechanical load could not be automatically inferred by so few motors. On the other hand, it might also be that the inference of the mechanical load based on only the transient current is simply not possible, because the motor type and/or its specific application influences the transient current much more than the type of the mechanical output.

\begin{figure} [htb]
	\centering
	\includegraphics[width=3.5in]{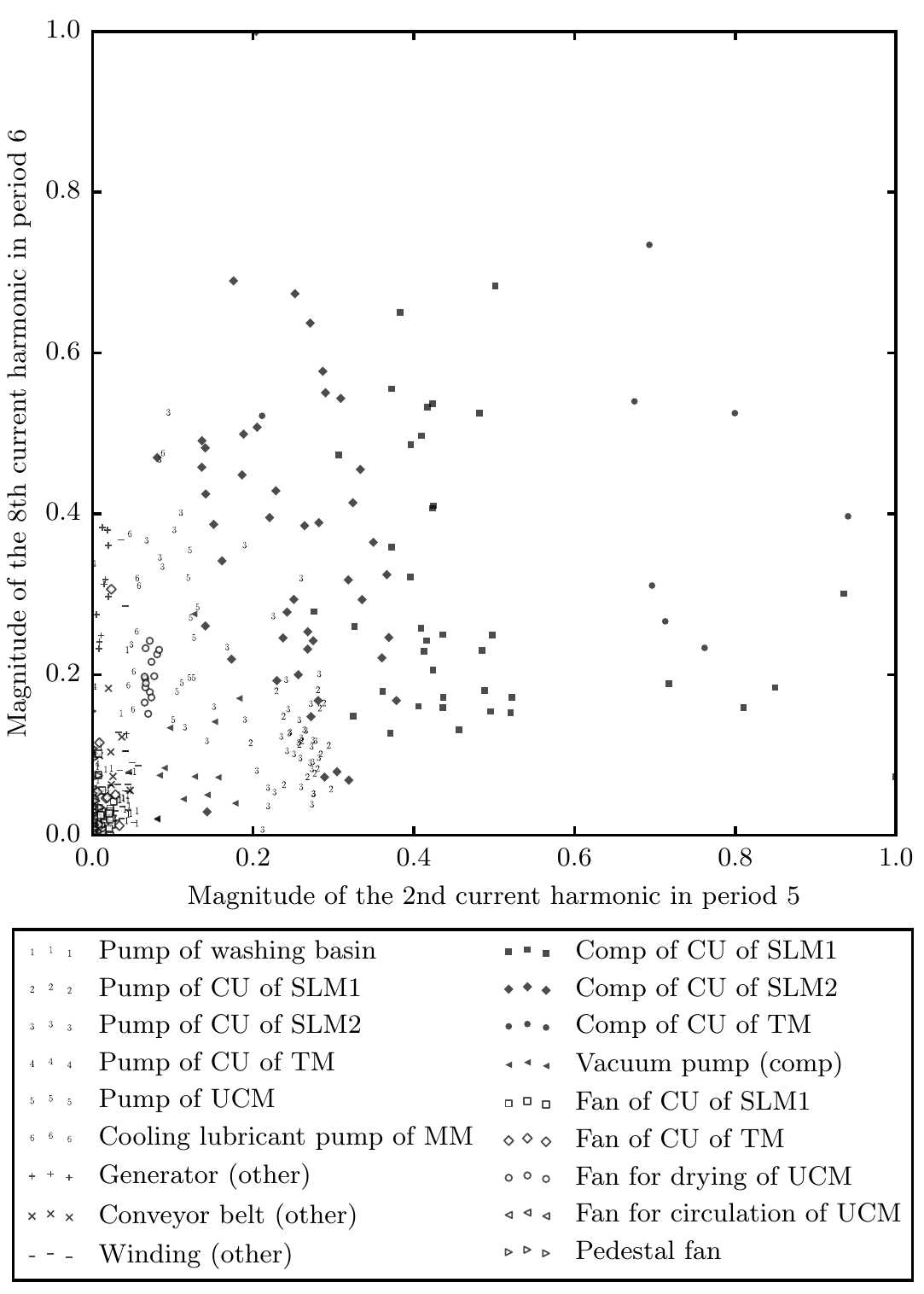}
	\caption{Scatter plot of all turn-on events. CU=cooling unit, SLM=selective laser sintering machine, TM= thermoform machine, MM=milling machine, UCM=ultrasonic cleaning machine.}
	\label{fig:result_load_scatter}
\end{figure}

The turn-on events can again be plotted in a scatter plot along the axis of the first two winning features. The second and third winning feature of the classifier using the linear kernel were the presence of inflection points in a particular half period. Since these are binary features, the resulting visualization is difficult to read. Instead, the first two winning features of the classifier using the radial basis function were used as axis, which represent the magnitude of the 2nd current harmonic in period 5 and the magnitude of the 8th current harmonic in period 6 (see Fig.~\ref{fig:result_load_scatter}). According to this graph, the turn-on evens from fans (white shapes) tend to exhibit fewer current harmonics than the turn-on events from compressors (black shapes). Turn-on events from pumps (numbers) fall somewhere in between. However, the class borders are more overlapping compared to Fig.~\ref{fig:result_motor_scatter}, which explains the lower classification performance.

\section{Summary}
\label{sec:summary}

We tested whether 18 different fixed-speed motors could be distinguished based upon their \emph{normalized}transient current. We found a classifier using 14 different features that can separate these motors with an f1-score of 97.7~\%. This can be considered a success compared to the f1-score of 5.6~\% from a randomly guessing classifier. Even when only five features can be used, a classifier was found that can distinguish the motors with an f1-score of approximately 80~\%.

In addition to that it was tested, whether the mechanical output type of a motor (i.e., a pump, a fan or a compressor) could be inferred by only analyzing the normalized transient current. Therefore, the classifier was trained with turn-on events from 3-5 motor for each mechanical output type, but it performed the prediction on turn-on events from motors which it has never seen before. In this way, the best classifier achieved f1-scores of approximately 70~\%. This is not a significant improvement compared to a randomly guessing classifier which yields an f1-score of 33.3~\%. This poor performance might be due to the low numbers of motors for each mechanical load type. However, it may also indicate that such an inference is simply not possible, because the motor type and/or the specific application influence the transient current characteristics more than the type of mechanical output.


%

\section*{Acknowledgment}
The authors would like to express their sincere thanks to the Freistaat Bayern for funding the
project "Green Factory Bavaria" in the framework of the future initiative "Aufbruch Bayern”

\appendices


\clearpage
\onecolumn
\section{Overview of existing papers concerning load identification}
\label{sec:attach_table_review}
\input{v6_table.tex}
\twocolumn
\clearpage

\newpage

\ifCLASSOPTIONcaptionsoff
  \newpage
\fi



\bibliographystyle{IEEEtran}


\bibliography{library_tim,library_table}

%
%
%

%

\begin{IEEEbiographynophoto}
{Christian Gebbe}
Christian Gebbe, born in 1987, studied applied physics (B.~Sc.) and electrical engineering and information technology (M.~Sc.) at the Technical University Munich. Since 2013 he has been researching how to extract more information from energy monitoring data of manufacturing companies at the Fraunhofer Research Institution for Casting, Composite and Processing Technology (IGCV) . His research focuses on using nonintrusive load monitoring to determine the electric energy demand of components of production machines.
\end{IEEEbiographynophoto}

\begin{IEEEbiographynophoto}
{Adil Bashir}
Adil Bashir, born in 1987, studied electronic engineering (B.Sc.) at the Sir Syed University of Engineering \& Technology. He educated himself further by studying Systems Engineering and Engineering Management (M.Sc.) at the South Westphalia University of Applied Sciences which he completed with a Master thesis at the Fraunhofer Research Institution for Casting, Composite and Processing Technology (IGCV).
\end{IEEEbiographynophoto}

\begin{IEEEbiographynophoto}
{Thomas Neuh\"auser}
Thomas Neuh\"auser, born in 1990, studied industrial engineering and management (M. Sc.) at the Technical University of Kaiserslautern. In 2018, he joined the Fraunhofer Research Institution for Casting, Composite and Processing Technology (IGCV) where he is working as a researcher and consultant on data analytics for production systems.
\end{IEEEbiographynophoto}


\vfill




\end{document}

%% file: v6_table.tex
\footnotesize

\tablehead{
	\hline
	Paper & Proposed categories & Proposed features & Proposed classifier & Classifier tested? & Input data & Input data & Loads in case study & Evaluation of case study \\
	\hline}
\begin{xtabular*}{\linewidth}{p{0.3cm}p{1.5cm}p{2cm}p{1.75cm}p{0.8cm}p{0.8cm}p{1cm}p{3.4cm}p{3cm}}
	
		\hline
		\cite{sultanem1991using} & One class per load & P,Q, shape and duration of transient, current harmonics & Iterative sorting using benchmark values and heuristics & yes & measured & aggregate & Appliances in occupied household for 24h (no details available) & Only qualitative: For appliances with simple operating cycles 95~\% accuracy. Issues were (1) duration of transient could not be determined reliably (2) appliances with several components combined in different manners (3) variable power appliances \\ \hline
		\cite{leeb1993development,leeb1993conjoint,norford1996non,khan1997multiprocessor} & Load classes, here: instant-start fluorescent lamps, rapid-start fluorescent lamps, induction motors & $P$, $Q$, $|I_3|$, $|I_3|^2$ during most varying part of transient (dubbed \mbox{v-section})  & transversal filter & yes & measured & per load and aggregate of two loads & Four loads: instant-start fluorescent lamp, rapid-start fluorescent lamp, two three-phase induction motors with ratings of 0.25~hp and 0.33~hp & Only qualitative: "[the classificator] performs remarkably well".  \\ \hline
		\cite{leeb1995transient} & Load classes specified in \cite{leeb1993development} plus personal computer & Time-varying Fourier coefficients of current & transversal filter & yes & measured & per load and aggregate of two loads & Same as in \cite{leeb1993development} & Only qualitative: "all events are correctly identified".  \\ \hline
		\cite{cole1998data} & not specified & Linear slope of active power & not specified & no, only approximation evaluated & measured & per load & A compressor of heat pump and a washing machine & Linear approximation of $p$ yields an error of less than 5~\%. \\ \hline
		\cite{lee2003electric} & Load classes similar to \cite{leeb1995transient} & Time-varying Fourier coefficients of transient current & pattern matching using least squares criterion in case of single events. Pattern matching using a more complex optimization in case of multi-event. & yes & Simulated and measured & per load and aggregate & In chapters 3-4 simulations of a few loads (e.g. pumps, fans). In chapter 8 a laundry room comprising nine dryers as well as a building comprising a variable speed drive. & In chapters 3-4 few quantitative results (mostly 100~\% accuracy in very limited tests). In chapter 8 no quantitative results. \\ \hline
		\cite{cox2006transient} & Load classes similar to \cite{leeb1995transient} but not specified in detail & Time-varying Fourier coefficients of voltage & fitting procedure performed as a series of decoupled estimation problems & yes & measured & aggregate & Four loads: 1~kW heater, 0.33 single-phase induction motor, three-way incandescent lamp, rapid-start fluorescent lamp & Only qualitative: "the complete system has been able to identify nearly all loads" \\ \hline
		\cite{srinivasan2006neural} & One class per load & Odd (complex) Fourier coefficients until 15th harmonic extracted from steady state current & 1. Multilayer perceptron neural network 2. Radial basis function neural network 3. Support vector machine with different kernels & yes & Measured (some aggregate data was simulated) & per load and aggregate. All possible combinations in training data. & Setup 1: monitor, fridge, fan, CPU, battery charger, television, light bulb, fluorescent lamp.
		
		Setup 2: CPUs, monitors, DC power supplies, notebook, phone charger, fluorescent lamp.
		
		Setup 3: motors, inverters, fluorescent lamps. & Accuracies between 53-100~\% depending on training set, generate noise and choice of classifier \\ \hline
		\cite{patel2007flick} & One class per load & Coefficients of fast Fourier transform from voltage in the frequency range between 0-50~kHz & Support vector machine (kernel not specified) & yes & measured & aggregate & Six different residental houses, whose appliances are not further specified. It is only reported that house 1 comprises 41 different appliances. Training data was collected by manually switching appliances. & Classification accuracies between 79-92~\% for different houses and different training sets \\ \hline
		\cite{chang2007load,yang2007design} & One class per load & $P$, $Q$, $V_{THD}$, $I_{THD}$, energy of transient $e$ defined as integration over $p_{inst}$ & Simple neural network trained by backpropagation  and learning vector quantization & yes & measured & per load and aggregate. All possible combinations in training data. & Five loads: three induction motors (95~hp, 140~hp, 300~hp), one induction motor controlled by a variable speed drive (95~hp), one load bank (R=4~$\Omega$) supplied by an uncontrolled three-phase bridge rectifier & For backpropagation: Accuracy between 95-100~\% already with only P,Q.
		
		For LVQ classifier; Additional use of $V_{THD}$, $I_{THD}$ and $e$ increase accuracy from 81-100~\% (individual loads) and  2-43~\% (load combinations) \\ \hline
		\cite{lam2007novel} & Load classes – different classification systems are discussed in paper & several features describing the shape of the normalized voltage-current-trajectory & Hierarchical clustering (not classification!) & yes (clustering only, not classification) & measured & per load & resistive appliances (electric kettle, heater, hair dryer), motor-driven appliances (vacuum cleaner, fan, blender), pump-operated appliances (refrigerator, dehumidifier), electronic appliances (computer, television, video recorder, energy-saving light bulb) as well as microwave oven and induction cooker & No accuracy, since only a clustering was performed, not a classification \\ \hline
		\cite{akbar2007modified} & Not clear: Either one class per load or or load classes similar to \cite{leeb1993conjoint} & Fourier coefficients of current & not specified & no, only analysis of data & measured & per load & compact fluorescent light, lamp with dimmer, fan with speed control & No  \\ \hline
		\cite{chang2008load} & same as in \cite{chang2007load} & P,Q, energy of transient & Simple neural network trained  by backpropagation  & yes & Case1 simulated
		
		Case2 measured & same as in \cite{chang2007load} & Case 1: two three-phase induction motors rated 2.6~hp and 4.7~hp and a R-L-linear load whose real and reactive power resembles the 4.7~hp motor.
		
		Case 2: 119~W dehumidifier, 590~W vacuum cleaner, R-L linear load whose real and reactive power resembles the 590~W vacuum cleaner & In both cases, accuracy for test data is approx. 40~\% using only P,Q and 100~\% using P,Q,e. \\ \hline
		\cite{katsukura2009life, kushiro2009non} & One class per load & Wavelet coefficients representing the value and phase of local peaks in current & unclear & yes & measured & per load and aggregate & A vacuum cleaner, an induction cooker, a microwave oven and an air conditioner & Accuracy of 100~\% if only one load is active and accuracies between 95-99~\% if multiple loads are active \\ \hline
		\cite{shaw2008nonintrusive} & Different load classes including incandescent lights and motors & Manually selected coefficients of time-varying Fourier series of transient current & pattern matching using least squares & yes & measured & aggregate & Case 1: Nine loads within a car comprising lights and motors.
		Case 2: Eight loads in an AC network comprising induction machines, a computer and different lights (instant start, compact fluorescent, rapid start, incandescent).
		Case 3: Loads within a naval vessel comprising induction motors driving pumps. & No quantitative results concerning transient classification are reported for any of the case studies. Instead, rather focus on condition monitoring. \\ \hline
		\cite{proper2009field} & same as \cite{lee2003electric} & same as \cite{lee2003electric} & same as \cite{lee2003electric} & yes & measured & aggregate & 1500 turn-on events of a naval waste collection and disposal system comprising mainly two water pumps and two discharge pumps & Accuracy of 95.9~\% \\ \hline
		\cite{chang2010load,chang2012new} & same as in \cite{chang2007load} & same as \cite{chang2008load} & same as \cite{chang2008load} & yes & Case1 simulated
		
		Case2 measured
		
		Case3,4 see \cite{chang2008load} & same as \cite{chang2007load} & Case 1: two three-phase induction motors rated 160~hp and 123~hp supplied by variable voltage drives and a load bank supplied by a controlled three-phase bridge rectifier (six pulse thyristor rectifier).
		
		Case 2: a one-phase induction motor rated 0.2~hp, a three phase induction motor rated 1~hp and a three phase R-L linear load.
		
		Case 3, 4 are equivalent to case 1, 2 in \cite{chang2008load}. & Case 1: accuracy for test data is already 100~\% using only P,Q.
		
		Case 2: In both cases, accuracy for test data is 95~\% using only P,Q and 100~\% using P,Q,e.
		
		Case 3, 4 are equivalent to case 1, 2 in \cite{chang2008load} \\ \hline
		\cite{shrestha2009dynamic} & same as \cite{lee2003electric} & same as \cite{lee2003electric} & same as \cite{lee2003electric} & yes & measured & aggregate & A shipboard power system comprising 15 static loads, 4 rectifier-type loads and 11 motors & No results reported concerning the load identification algorithm. Only a few qualitative statements such as “NILM succesfully detected these changes” \\ \hline
		\cite{ruzzelli2010real} & One class per appliance type (e.g. washing machine) & real power, power factor, peak current, RMS current, peak voltage, RMS voltage & neural network & yes & measured & aggregate & kettle, heater, microwave, fridge & Accuracy of 95~\% in scenario without heater, Accuracy of 84~\% in scenario with heater \\ \hline
		\cite{chang2012non} & same as in \cite{chang2007load} & same as in \cite{chang2007load} + duration of transient $t_{TR}$ determined by either DWT or STFT coefficients & same as in \cite{chang2008load} & yes & same as in \cite{chang2010load} & same as in \cite{chang2007load} & Case 1: same as case1 in \cite{chang2010load}.
		Case 2: same as case1 in \cite{chang2008load}.
		Case 3: same as case2 in \cite{chang2010load}.
		Case 4: same as case2 in \cite{chang2010load}. & Case 1: Accuracy is 100~\%, even when using only P,Q.
		
		Case 2: Accuracy increases from 53~\% (only P,Q) to 92~\% (only e, $t_{TR}$).
		
		Case 3: Accuarcy increases from 92~\% (only P,Q) to 97~\% (only e, $t_{TR}$).
		
		Case 4: Accuarcy increases from 48~\% (only P,Q) to 83~\% (only e, $t_{TR}$). \\ \hline
		\cite{lin2012application} & One class per load & Crest Factor and peak value of current & Neuro fuzzy pattern recognition combined with fuzzy c-means clustering & yes & measured & per load and aggregate & A fan, a fluorescent light and a radio & Accuracies between 86-100~\% \\ \hline
		\cite{tsai2012modern} & One class per load & peak, average and rms of current as well as duration of transient & K-nearest neighbor and  neural network & yes & measured & per load and aggregate & A fan, a fluorescent light, a radio and a microwave oven & Accuracies between 84-100~\% \\ \hline
		\cite{bouhouras2012load} & One class per load & seven Shanon entropy coefficients calculated using FFT extracted from current in steady state & nearest neighbor using Euclidian distance & yes & measured & per load & 16 loads: coffee machine, hair dryer (full and half power), halogen light, electric heater (full power, half power and standby), home theater, air conditioner (with and without inverter), iron, laptop, LED light, luminaire, refrigerator, and washing machine & Accuracies of 100~\% \\ \hline
		\cite{chang2013particle} & same as in \cite{chang2007load} & (rescaled) real and reactive power & neural network trained with particle swarm optimization method & yes & Case1 simulated
		
		Case2 measured
		
		Case3 measured & same as in \cite{chang2007load} & Case 1: same as case1 in \cite{chang2010load}
		
		Case 2: same as case2 in \cite{chang2010load}
		
		Case 3: A 910W air-conditioner, a 100W television, a 1200W vacuum cleaner, a 780W hair dryer and a 168W desktop computer & Case 1: Accuracy of 99.2~\% on validation data
		
		Case 2: Accuracy of 92.9~\% on validation data
		
		Case 3: Accuracy of 96.7~\% on validation data \\ \hline
		\cite{fernandes2013load} & One class per load & Amplitude of first, third, fifth, seventh, eigth and nineth current harmonics & one neural network for each load & yes & measured & per load (?) & personal computer, monitor, compact fluorescent light, fluorescent light, incandescent light, ventilator & Mean precision of 98.3~\% \\ \hline
		\cite{paradiso2013ann} & One class per  appliance type (e.g. washing machine, dishwaser, microwave oven) & Ten custom features extracted from active power signal & neural network & yes & measured & per load & Washing machine, refrigerator, dishwasher, smart TV, iron, microwave oven, lighting, coffee machine & Average accuracy of 95.3~\% (test case 1) \\ \hline
		\cite{he2013incorporating} & One class per load & fundamental phase angle, third component harmonic, fundamental ratio, fifth component harmonic fundamental ratio, transient impedance & Self-organized map & yes & measured & per load (?) & space heater, furnace, water heater, lightning, fan, TV, printer, desktop PC, projector, monitor & Accuracies between 77-99~\% \\ \hline
		\cite{chang2014power} & same as in \cite{chang2007load} & Power spectrum of the wavelet transform coefficients from current & neural network trained by backpropagation & yes & Simulated and measured & same as in \cite{chang2007load} & Case 1: same as case1 in \cite{chang2010load}
		
		Case 2: subset of case 3 in \cite{chang2013particle} (only television, vacuum cleaner and hair dryer)
		
		Case 3: same as case1 in \cite{chang2008load}
		
		Case 4: 100W television, 80W lamp, 100W fan
		
		Case 5: same as case3 with simultaneous turn-on events
		
		Case 6: same as case3 for turn-off events & Depending on training data:
		
		Case 1: Accuracy of 97-100~\%
		Case 2: Accuracy of 88-97~\%
		Case 3: Accuracy of 99-100~\%
		Case 4: Accuracy of 86-97~\%
		Case 5: Accuracy of 0-100~\%
		Case 6: Accuracy of 90-100~\% \\ \hline
		\cite{lin2014development} & One class per load & time frequency based S-transform (similar to short time Fourier transform) & ant colonization algorithm search algorithm & yes & measured & aggregate & A vacuum cleaner, an electric boiler, a microwave oven and a hair dryer & Accuracy of 79~\% \\ \hline
		\cite{bier2014disaggregation} & One class per load & real, reactive and apparent power as well as short time Fourier transform coefficients of transient current & neural network & yes & measured & per load & water heater, freezer, hand held mixer, two different hair-dryer, mixer, incandescent lamp, heater oven, energy saving lamp, radio & Accuracies between 0-100~\% depending on appliance, selected feature and sample length \\ \hline
		\cite{vsira2015system} & see \cite{bouhouras2012load} & see \cite{bouhouras2012load} & see \cite{bouhouras2012load} & yes & simulated (based on measurements from \cite{bouhouras2012load}) & see \cite{bouhouras2012load} & see \cite{bouhouras2012load} & Accuracies between 6-100~\%, mean accuracy between 77-99~\% depending on type of input 

\end{xtabular*}

%% file: v6.bbl
\begin{thebibliography}{10}
\providecommand{\url}[1]{#1}
\csname url@samestyle\endcsname
\providecommand{\newblock}{\relax}
\providecommand{\bibinfo}[2]{#2}
\providecommand{\BIBentrySTDinterwordspacing}{\spaceskip=0pt\relax}
\providecommand{\BIBentryALTinterwordstretchfactor}{4}
\providecommand{\BIBentryALTinterwordspacing}{\spaceskip=\fontdimen2\font plus
\BIBentryALTinterwordstretchfactor\fontdimen3\font minus
  \fontdimen4\font\relax}
\providecommand{\BIBforeignlanguage}[2]{{%
\expandafter\ifx\csname l@#1\endcsname\relax
\typeout{** WARNING: IEEEtran.bst: No hyphenation pattern has been}%
\typeout{** loaded for the language `#1'. Using the pattern for}%
\typeout{** the default language instead.}%
\else
\language=\csname l@#1\endcsname
\fi
#2}}
\providecommand{\BIBdecl}{\relax}
\BIBdecl

\bibitem{zoha2012non}
A.~Zoha, A.~Gluhak, M.~Imran, and S.~Rajasegarar, ``Non-intrusive load
  monitoring approaches for disaggregated energy sensing: A survey,''
  \emph{Sensors}, vol.~12, no.~12, pp. 16\,838--16\,866, 2012.

\bibitem{du2010review}
Y.~Du, L.~Du, B.~Lu, R.~Harley, and T.~Habetler, ``A review of identification
  and monitoring methods for electric loads in commercial and residential
  buildings,'' in \emph{2010 IEEE Energy Conversion Congress and
  Exposition}.\hskip 1em plus 0.5em minus 0.4em\relax IEEE, 2010, pp.
  4527--4533.

\bibitem{sultanem1991using}
F.~Sultanem, ``Using appliance signatures for monitoring residential loads at
  meter panel level,'' \emph{IEEE Transactions on Power Delivery}, vol.~6,
  no.~4, pp. 1380--1385, 1991.

\bibitem{leeb1993development}
S.~Leeb, J.~L. Kirtley~Jr, M.~S. LeVan, and J.~P. Sweeney, ``Development and
  validation of a transient event detector,'' \emph{AMP Journal of Technology},
  vol.~3, pp. 69--74, 1993.

\bibitem{leeb1993conjoint}
S.~B. Leeb, ``A conjoint pattern recognition approach to nonintrusive load
  monitoring,'' Ph.D. dissertation, Massachusetts Institute of Technology,
  1993.

\bibitem{norford1996non}
L.~K. Norford and S.~B. Leeb, ``Non-intrusive electrical load monitoring in
  commercial buildings based on steady-state and transient load-detection
  algorithms,'' \emph{Energy and Buildings}, vol.~24, no.~1, pp. 51--64, 1996.

\bibitem{khan1997multiprocessor}
U.~A. Khan, S.~B. Leeb, and M.~C. Lee, ``A multiprocessor for transient event
  detection,'' \emph{IEEE Transactions on Power Delivery}, vol.~12, no.~1, pp.
  51--60, 1997.

\bibitem{leeb1995transient}
S.~B. Leeb, S.~R. Shaw, and J.~L. Kirtley, ``Transient event detection in
  spectral envelope estimates for nonintrusive load monitoring,'' \emph{IEEE
  Transactions on Power Delivery}, vol.~10, no.~3, pp. 1200--1210, 1995.

\bibitem{cole1998data}
A.~I. Cole and A.~Albicki, ``Data extraction for effective non-intrusive
  identification of residential power loads,'' in \emph{IMTC/98 Conference
  Proceedings. IEEE Instrumentation and Measurement Technology Conference.
  Where Instrumentation is Going (Cat. No. 98CH36222)}, vol.~2.\hskip 1em plus
  0.5em minus 0.4em\relax IEEE, 1998, pp. 812--815.

\bibitem{lee2003electric}
K.~D. Lee, ``Electric load information system based on non-intrusive power
  monitoring,'' Ph.D. dissertation, Massachusetts Institute of Technology,
  2003.

\bibitem{cox2006transient}
R.~Cox, S.~B. Leeb, S.~R. Shaw, and L.~K. Norford, ``Transient event detection
  for nonintrusive load monitoring and demand side management using voltage
  distortion,'' in \emph{Twenty-First Annual IEEE Applied Power Electronics
  Conference and Exposition, 2006. APEC'06.}\hskip 1em plus 0.5em minus
  0.4em\relax IEEE, 2006, pp. 7--pp.

\bibitem{srinivasan2006neural}
D.~Srinivasan, W.~Ng, and A.~Liew, ``Neural-network-based signature recognition
  for harmonic source identification,'' \emph{IEEE Transactions on Power
  Delivery}, vol.~21, no.~1, pp. 398--405, 2006.

\bibitem{patel2007flick}
S.~N. Patel, T.~Robertson, J.~A. Kientz, M.~S. Reynolds, and G.~D. Abowd, ``At
  the flick of a switch: Detecting and classifying unique electrical events on
  the residential power line (nominated for the best paper award),'' in
  \emph{International Conference on Ubiquitous Computing}.\hskip 1em plus 0.5em
  minus 0.4em\relax Springer, 2007, pp. 271--288.

\bibitem{chang2007load}
H.-H. Chang, H.-T. Yang, and C.-L. Lin, ``Load identification in neural
  networks for a non-intrusive monitoring of industrial electrical loads,'' in
  \emph{International Conference on Computer Supported Cooperative Work in
  Design}.\hskip 1em plus 0.5em minus 0.4em\relax Springer, 2007, pp. 664--674.

\bibitem{yang2007design}
H.-T. Yang, H.-H. Chang, and C.-L. Lin, ``Design a neural network for features
  selection in non-intrusive monitoring of industrial electrical loads,'' in
  \emph{2007 11th International Conference on Computer Supported Cooperative
  Work in Design}.\hskip 1em plus 0.5em minus 0.4em\relax IEEE, 2007, pp.
  1022--1027.

\bibitem{lam2007novel}
H.~Y. Lam, G.~Fung, and W.~Lee, ``A novel method to construct taxonomy
  electrical appliances based on load signaturesof,'' \emph{IEEE Transactions
  on Consumer Electronics}, vol.~53, no.~2, pp. 653--660, 2007.

\bibitem{akbar2007modified}
M.~Akbar and Z.~A. Khan, ``Modified nonintrusive appliance load monitoring for
  nonlinear devices,'' in \emph{2007 IEEE International Multitopic
  Conference}.\hskip 1em plus 0.5em minus 0.4em\relax IEEE, 2007, pp. 1--5.

\bibitem{chang2008load}
H.-H. Chang, C.-L. Lin, and H.-T. Yang, ``Load recognition for different loads
  with the same real power and reactive power in a non-intrusive
  load-monitoring system,'' in \emph{2008 12th International Conference on
  Computer Supported Cooperative Work in Design}.\hskip 1em plus 0.5em minus
  0.4em\relax IEEE, 2008, pp. 1122--1127.

\bibitem{katsukura2009life}
M.~Katsukura, M.~Nakata, Y.~Itou, and N.~Kushiro, ``Life pattern sensor with
  non-intrusive appliance monitoring,'' in \emph{2009 Digest of Technical
  Papers International Conference on Consumer Electronics}.\hskip 1em plus
  0.5em minus 0.4em\relax IEEE, 2009, pp. 1--2.

\bibitem{kushiro2009non}
N.~Kushiro, M.~Katsukura, M.~Nakata, and Y.~Ito, ``Non-intrusive human behavior
  monitoring sensor for health care system,'' in \emph{Symposium on Human
  Interface}.\hskip 1em plus 0.5em minus 0.4em\relax Springer, 2009, pp.
  549--558.

\bibitem{shaw2008nonintrusive}
S.~R. Shaw, S.~B. Leeb, L.~K. Norford, and R.~W. Cox, ``Nonintrusive load
  monitoring and diagnostics in power systems,'' \emph{IEEE Transactions on
  Instrumentation and Measurement}, vol.~57, no.~7, pp. 1445--1454, 2008.

\bibitem{proper2009field}
E.~Proper, R.~W. Cox, S.~B. Leeb, K.~Douglas, J.~Paris, W.~Wichakool, E.~L.
  Foulks, R.~Jones, P.~Branch, A.~Fuller \emph{et~al.}, ``Field demonstration
  of a real-time non-intrusive monitoring system for condition-based
  maintenance,'' Massachusetts Institute of Technology. Sea Grant College
  Program, Tech. Rep., 2009.

\bibitem{chang2010load}
H.-H. Chang, C.-L. Lin, and J.-K. Lee, ``Load identification in nonintrusive
  load monitoring using steady-state and turn-on transient energy algorithms,''
  in \emph{The 2010 14th International Conference on Computer Supported
  Cooperative Work in Design}.\hskip 1em plus 0.5em minus 0.4em\relax IEEE,
  2010, pp. 27--32.

\bibitem{chang2012new}
H.-H. Chang, K.-L. Chen, Y.-P. Tsai, and W.-J. Lee, ``A new measurement method
  for power signatures of nonintrusive demand monitoring and load
  identification,'' \emph{IEEE Transactions on Industry Applications}, vol.~48,
  no.~2, pp. 764--771, 2012.

\bibitem{shrestha2009dynamic}
A.~Shrestha, E.~L. Foulks, and R.~W. Cox, ``Dynamic load shedding for shipboard
  power systems using the non-intrusive load monitor,'' in \emph{2009 IEEE
  Electric Ship Technologies Symposium}.\hskip 1em plus 0.5em minus 0.4em\relax
  IEEE, 2009, pp. 412--419.

\bibitem{ruzzelli2010real}
A.~G. {Ruzzelli}, C.~{Nicolas}, A.~{Schoofs}, and G.~M.~P. {O'Hare},
  ``Real-time recognition and profiling of appliances through a single
  electricity sensor,'' in \emph{2010 7th Annual IEEE Communications Society
  Conference on Sensor, Mesh and Ad Hoc Communications and Networks (SECON)},
  June 2010, pp. 1--9.

\bibitem{chang2012non}
H.-H. Chang, ``Non-intrusive demand monitoring and load identification for
  energy management systems based on transient feature analyses,''
  \emph{Energies}, vol.~5, no.~11, pp. 4569--4589, 2012.

\bibitem{lin2012application}
Y.-H. Lin and M.-S. Tsai, ``Application of neuro-fuzzy pattern recognition for
  non-intrusive appliance load monitoring in electricity energy conservation,''
  in \emph{2012 IEEE International Conference on Fuzzy Systems}.\hskip 1em plus
  0.5em minus 0.4em\relax IEEE, 2012, pp. 1--7.

\bibitem{tsai2012modern}
M.-S. Tsai and Y.-H. Lin, ``Modern development of an adaptive non-intrusive
  appliance load monitoring system in electricity energy conservation,''
  \emph{Applied Energy}, vol.~96, pp. 55--73, 2012.

\bibitem{bouhouras2012load}
A.~Bouhouras, A.~Milioudis, G.~Andreou, and D.~Labridis, ``Load signatures
  improvement through the determination of a spectral distribution coefficient
  for load identification,'' in \emph{2012 9th International Conference on the
  European Energy Market}.\hskip 1em plus 0.5em minus 0.4em\relax IEEE, 2012,
  pp. 1--6.

\bibitem{chang2013particle}
H.-H. Chang, L.-S. Lin, N.~Chen, and W.-J. Lee,
  ``Particle-swarm-optimization-based nonintrusive demand monitoring and load
  identification in smart meters,'' \emph{IEEE Transactions on Industry
  Applications}, vol.~49, no.~5, pp. 2229--2236, 2013.

\bibitem{fernandes2013load}
R.~A.~S. {Fernandes}, I.~N. {da Silva}, and M.~{Oleskovicz}, ``Load profile
  identification interface for consumer online monitoring purposes in smart
  grids,'' \emph{IEEE Transactions on Industrial Informatics}, vol.~9, no.~3,
  pp. 1507--1517, Aug 2013.

\bibitem{paradiso2013ann}
F.~{Paradiso}, F.~{Paganelli}, A.~{Luchetta}, D.~{Giuli}, and
  P.~{Castrogiovanni}, ``Ann-based appliance recognition from low-frequency
  energy monitoring data,'' in \emph{2013 IEEE 14th International Symposium on
  "A World of Wireless, Mobile and Multimedia Networks" (WoWMoM)}, June 2013,
  pp. 1--6.

\bibitem{he2013incorporating}
D.~{He}, W.~{Lin}, N.~{Liu}, R.~G. {Harley}, and T.~G. {Habetler},
  ``Incorporating non-intrusive load monitoring into building level demand
  response,'' \emph{IEEE Transactions on Smart Grid}, vol.~4, no.~4, pp.
  1870--1877, Dec 2013.

\bibitem{chang2014power}
H.-H. Chang, K.-L. Lian, Y.-C. Su, and W.-J. Lee, ``Power-spectrum-based
  wavelet transform for nonintrusive demand monitoring and load
  identification,'' \emph{IEEE Transactions on Industry Applications}, vol.~50,
  no.~3, pp. 2081--2089, 2014.

\bibitem{lin2014development}
Y.-H. Lin and M.-S. Tsai, ``Development of an improved time--frequency
  analysis-based nonintrusive load monitor for load demand identification,''
  \emph{IEEE Transactions on Instrumentation and Measurement}, vol.~63, no.~6,
  pp. 1470--1483, 2014.

\bibitem{bier2014disaggregation}
T.~Bier, ``Disaggregation of electrical appliances using non-intrusive load
  monitoring,'' Ph.D. dissertation, Universit{\'e} de Haute Alsace-Mulhouse,
  2014.

\bibitem{vsira2015system}
M.~{\v{S}}{\'\i}ra and V.~N. Zachovalov{\'a}, ``System for calibration of
  nonintrusive load meters with load identification ability,'' \emph{IEEE
  Transactions on Instrumentation and Measurement}, vol.~64, no.~6, pp.
  1350--1354, 2015.

\bibitem{zeifman2011nonintrusive}
M.~Zeifman and K.~Roth, ``Nonintrusive appliance load monitoring: Review and
  outlook,'' \emph{IEEE transactions on Consumer Electronics}, vol.~57, no.~1,
  pp. 76--84, 2011.

\bibitem{marchildon2014pumps}
K.~Marchildon and D.~Mody, ``Pumps, fans, blowers, and compressors,''
  \emph{Mechanical Engineers' Handbook}, pp. 1--44, 2014.

\bibitem{gupta2010electrisense}
S.~Gupta, M.~S. Reynolds, and S.~N. Patel, ``Electrisense: single-point sensing
  using emi for electrical event detection and classification in the home,'' in
  \emph{Proceedings of the 12th ACM international conference on Ubiquitous
  computing}.\hskip 1em plus 0.5em minus 0.4em\relax ACM, 2010, pp. 139--148.

\bibitem{suzuki2008nonintrusive}
K.~Suzuki, S.~Inagaki, T.~Suzuki, H.~Nakamura, and K.~Ito, ``Nonintrusive
  appliance load monitoring based on integer programming,'' in \emph{2008 SICE
  Annual Conference}.\hskip 1em plus 0.5em minus 0.4em\relax IEEE, 2008, pp.
  2742--2747.

\bibitem{egarter2013evolving}
D.~Egarter, A.~Sobe, and W.~Elmenreich, ``Evolving non-intrusive load
  monitoring,'' in \emph{European Conference on the Applications of
  Evolutionary Computation}.\hskip 1em plus 0.5em minus 0.4em\relax Springer,
  2013, pp. 182--191.

\bibitem{kim2011unsupervised}
H.~Kim, M.~Marwah, M.~Arlitt, G.~Lyon, and J.~Han, ``Unsupervised
  disaggregation of low frequency power measurements,'' in \emph{Proceedings of
  the 2011 SIAM international conference on data mining}.\hskip 1em plus 0.5em
  minus 0.4em\relax SIAM, 2011, pp. 747--758.

\bibitem{parson2014unsupervised}
O.~Parson, S.~Ghosh, M.~Weal, and A.~Rogers, ``An unsupervised training method
  for non-intrusive appliance load monitoring,'' \emph{Artificial
  Intelligence}, vol. 217, pp. 1--19, 2014.

\bibitem{refaeilzadeh2009cross}
P.~Refaeilzadeh, L.~Tang, and H.~Liu, ``Cross-validation,'' in
  \emph{Encyclopedia of database systems}.\hskip 1em plus 0.5em minus
  0.4em\relax Springer, 2009, pp. 532--538.

\bibitem{gebbe2017feature}
C.~Gebbe, C.~Tran, F.~Lingenfelser, J.~Glasschr{\"o}der, and G.~Reinhart,
  ``Feature extraction and classification of the electric current signal of an
  induction motor for condition monitoring purposes.'' in \emph{Applied
  Mechanics \& Materials}, vol. 856.\hskip 1em plus 0.5em minus 0.4em\relax
  Trans Tech Publications, 2017, pp. 244--251.

\bibitem{maglogiannis2007emerging}
I.~G. Maglogiannis, \emph{Emerging artificial intelligence applications in
  computer engineering: real word ai systems with applications in ehealth, hci,
  information retrieval and pervasive technologies}.\hskip 1em plus 0.5em minus
  0.4em\relax Ios Press, 2007, vol. 160.

\bibitem{pedregosa2011scikit}
F.~Pedregosa, G.~Varoquaux, A.~Gramfort, V.~Michel, B.~Thirion, O.~Grisel,
  M.~Blondel, P.~Prettenhofer, R.~Weiss, V.~Dubourg \emph{et~al.},
  ``Scikit-learn: Machine learning in python,'' \emph{Journal of machine
  learning research}, vol.~12, no. Oct, pp. 2825--2830, 2011.

\end{thebibliography}
